\documentclass[aps,prl,twocolumn,letterpaper,superscriptaddress,showpacs]{revtex4-1}

\usepackage{graphicx}
\usepackage{dcolumn}
\usepackage{bm}
\usepackage{verbatim}
\usepackage{subfigure}
\usepackage{multirow}
\usepackage{setspace}
\usepackage{physics}
\usepackage{xcolor}
\usepackage{siunitx}
\usepackage{CJK}
\usepackage{mathptmx}
\usepackage[colorlinks,linkcolor=blue,anchorcolor=blue, citecolor=blue,urlcolor=blue,]{hyperref}

\begin{document}

\begin{CJK*}{UTF8}{bsmi}
\title{2D ferroelectricity accompanying antiferro-orbital order in semi-metallic WTe$_{2}$}

\author{Fangyuan Gu (\CJKfamily{gbsn}顾方圆)}
\altaffiliation{These authors contributed equally.}
\affiliation{Tsung-Dao Lee Institute and School of Physics and Astronomy, Shanghai Jiao Tong University, Shanghai 200240, China}
\affiliation{School of Materials Science and Engineering, Shanghai Jiao Tong University, Shanghai 200240, China}

\author{Ruoshi Jiang (\CJKfamily{gbsn}姜若诗)}
\altaffiliation{These authors contributed equally.}
\affiliation{Tsung-Dao Lee Institute and School of Physics and Astronomy, Shanghai Jiao Tong University, Shanghai 200240, China}
\affiliation{Department of Materials Science and Metallurgy, University of Cambridge, Cambridge CB3 0FS, United Kingdom}

\author{Wei Ku (\CJKfamily{bsmi}顧威)}
\altaffiliation{corresponding email: weiku@sjtu.edu.cn}
\affiliation{Tsung-Dao Lee Institute and School of Physics and Astronomy, Shanghai Jiao Tong University, Shanghai 200240, China}
\affiliation{Key Laboratory of Artificial Structures and Quantum Control (Ministry of Education), Shanghai 200240, China}
\affiliation{Shanghai Branch, Hefei National Laboratory, Shanghai 201315, People's Republic of China \vspace{4pt}}

\date{\today}

\begin{abstract}
The first switchable electric polarization in \textit{metals} was recently discovered in bilayer and trilayer WTe$_2$.
Strangely, despite the tininess of the ordered polarization, the ferroelectricity survives up to $\sim$350 K, rendering the mechanism of such ferroelectricity challenging for standard understandings.
Here, via a density-functional-based multi-energy-scale analysis of the system's broken symmetries, we identify a weak out-of-plane ferroelectricity accompanying a strong in-plane antiferro-orbital order.
This unusual low-energy correlation, which emerges from an antiferroelectric structure formed at much higher energy, naturally explains the above puzzling observation.
This result reveals an unprecedented paradigm of \textit{electronic} ferroelectricity generally applicable to 2D polar metals with ultrafast-switchable polarization ideal for the next-generation non-volatile memory and other devices.
\end{abstract}
\maketitle
\end{CJK*}


Two-dimensional (2D) semi-metallic bilayer and trilayer WTe$_2$ are recently shown~\cite{Fei_nature2018} to host an exciting (and the first known) switchable electric polarization in metals.
The same was then found also in thin films of WTe$_2$ as well~\cite{Sharma_sciadv2019}. 
In contrast to the efficient screening of macroscopic polarization~\cite{Jirapa_FE_2009, Ivanchik_FE_1993, Gu_2023npj, Gu_PRB_2024} in conventional polar metals, \textit{metallic} WTe$_2$ retains a measurable macroscopic polarization under ambient conditions.
Such a combination of planar metallicity and perpendicular polarization in a capacitor-like geometry opens up promising applications in non-volatile memory and nanoelectronic devices.
Correspondingly, significant research activities have been carried out on various interesting physical properties of this material~\cite{Ali_nat2014, Jia_Natphy2022, Ma_Nat2019, Wang_npjcomp2019, Yang_JoP2022, Sie_Nat2019, Ji_ACSnano2021, Soloyanov_Nat2015}.

Concerning the essential macroscopic electric polarization in the metallic WTe$_2$, an obvious puzzle is the surprisingly high ferroelectric (FE) transition temperature $T_c \sim 350$ K in this material, given the tininess of the polarization density, $\sim 2\times 10^{11}\ e/$cm$^2$~\cite{Fei_nature2018}.
Specifically, since this polarization density is about \textit{three orders} of magnitude smaller than that in typical FE perovskites, for example BaTiO$_3$~\cite{Dawber_RevModPhys2005, Gu_PRM_2021} with $T_c \sim 400$ K, a na\"{i}ve estimation would expect a similar three orders of magnitude reduction in the transition temperature, rather than the same order at 350 K~\cite{Zhang_Natreview2023}.
It is therefore timely to elucidate the essential microscopic mechanism for the quantum states of matter that \textit{simultaneously} accounts for the observed high transition temperature under such a small polarization density.

Currently, the main proposals for the ferroelectricity in this material mostly rely on broken inversion symmetry in the atomic lattice.
Particularly, a popular proposal is based on the atomic position sliding between the adjacent layers, which in principle breaks the out-of-plane inversion symmetry and produces a very small electric dipole.
However, the estimated energy barrier ($\sim$0.3 meV) between the two symmetric sliding structures is too small to account for the high-transition temperature~\cite{Liu_Nanoscale2019, Yang_Jpcl2018, Ali_nat2014}.  On the other hand, transition through an unrealistic pathway involving a fully symmetric $1T$ atomic structure~\cite{Ji_ACSnano2021, Lee_SciRep2015, Tao_PhysRevB2020} would require 300-700 meV~\cite{Sharma_sciadv2019} per unit cell, more than an order of magnitude larger than the transition temperature.
Such an atomic position-based picture evidently does not even have the correct energy scale to offer a satisfactory resolution to the puzzling physical behavior of this material.

In fact, such atomic position-based pictures are inconsistent with many experimental observations.
For example, to date, there is no direct experimental evidence of structural phase transition at 350 K to recover the out-of-plane inversion symmetry in the atomic structure in either bulk or few-layer samples.
(In other words, the atomic structure is already stuck in one of the interlayer sliding modes above 350 K.)
Furthermore, in bulk samples even the monoclinic ($1T'$) bulk lattice structure above 565 K~\cite{Tao_PhysRevB2020, Zhou_AIPadv2016} still does \textit{not} recover this inversion symmetry between layers either.
(The electric polarization is merely compensated due to the alteration of dipole directions across bilayers.)
As a matter of fact, the presence of pre-existing broken inversion symmetry in the low-temperature FE phase is evident from the lack of symmetry in the hysteresis curve of conductance measurements~\cite{Fei_nature2018, Sharma_sciadv2019, Xiao2020_natphy}, including the observed switching field.
Therefore, the 350 K phase transition in bilayer samples, as well as the rapid development of electric polarization below, must involve physics \textit{additional to} the already broken out-of-plane inversion symmetry in the atomic structure.
Of course, such atomic position-based pictures are inapplicable to monolayer FE materials~\cite{Yuan_NatComm2019}.

More insightful clues on the dominant mechanism for the low-temperature phase are revealed in the response of the phase to weak applied electric field along the out-of-plane direction.
First, the required switching field $2\delta \mathrm{V}$ of the macroscopic polarization is found to be very small (with interlayer potential $\sim 40 \ \mathrm{mV}$)~\cite{Fei_nature2018}.
Such a small potential is insufficient to induce a global interlayer sliding of heavy atoms such as W and Te.
It is, on the other hand, comparable to the intrinsic electric potential from the observed polarization density, $2\times 10^{11}\ e/$cm$^2$, and thus well within the required scale to reshape the spatial distribution of the much lighter electrons.
Second, an unusually strong sensitivity to the applied field is observed in the conductance~\cite{Fei_nature2018}.
Consistent with the small switching field, this observation suggests the important involvement of electronic degrees of freedom in this puzzling low-temperature FE phase below 350 K.

\begin{figure}
\centering
\includegraphics[width=\columnwidth]{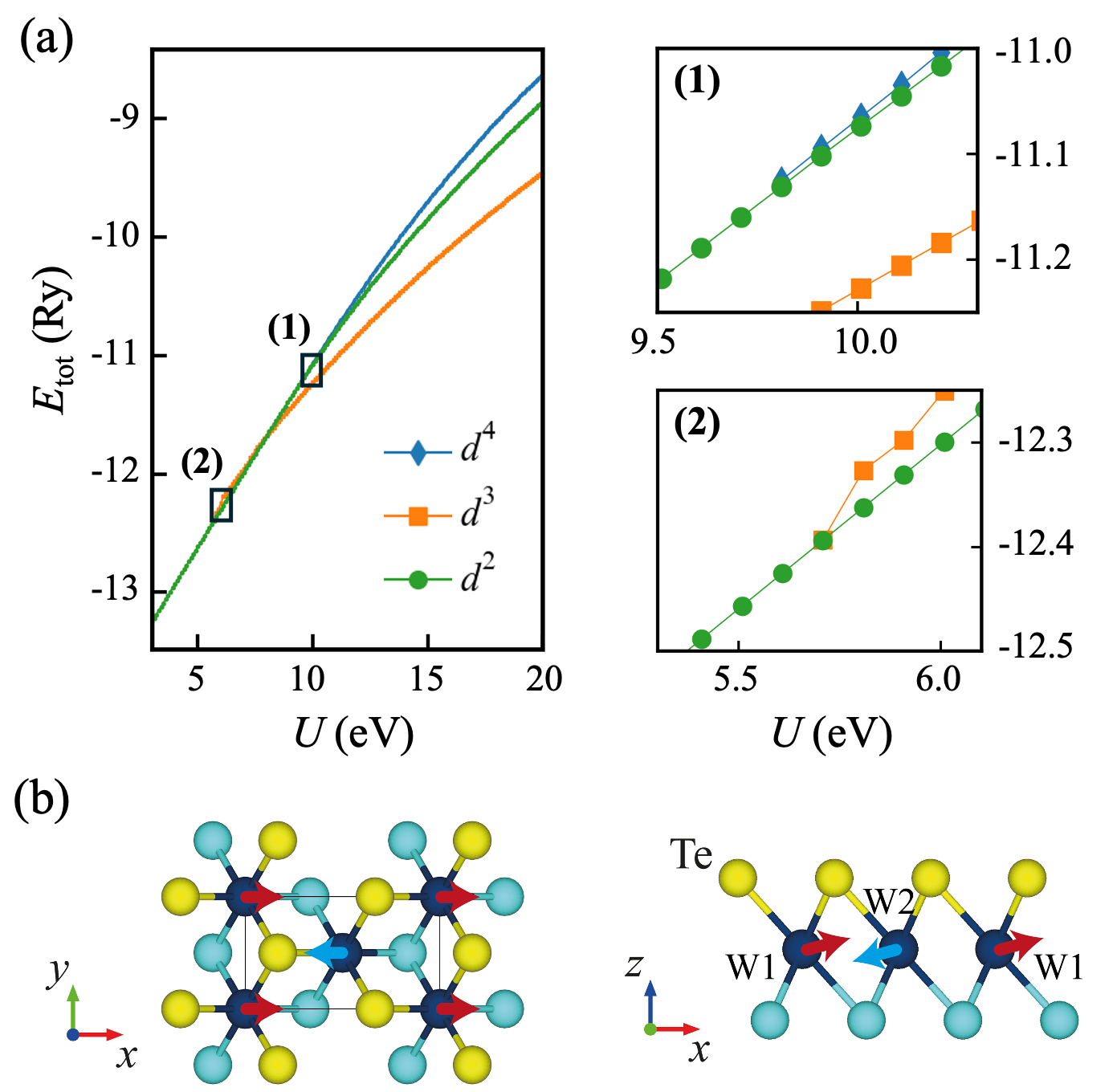}
\vspace{-0.5cm}
\caption{Effective $d^2$ valence of W and the resulting antiferroelectric (AFE) lattice distortion.
(a) Adiabatic connection of total energy, $E_\mathrm{tot}$, (and other aspects of the electronic structure) to well-defined ionic $d^2$ configurations at larger $U$ indicates that the real material has an effective $d^2$ valence, or equivalently W$^{4+}$Te$^{2-}_2$.
Insets (1) and (2) show the destabilization of $d^4$ and $d^3$ configurations below $U\sim 9.8$ and 5.7 eV, respectively.
(b) Exaggerated illustration (via red and blue arrows) of the AFE $T_d$ lattice distortion from the more symmetric $1T$ structure under such an effective W$^{4+}$Te$^{2-}_2$ ions.
}
\vspace{-0.5cm}
\label{fig:d2_AFEO}
\end{figure}

Here, motivated by these overwhelming experimental evidences, we investigate the electronic structure of WTe$_2$ to seek potential electronic mechanisms of the low-temperature FE phase. 
Through a density-functional-based multi-energy-scale analysis of the system's broken symmetries, we identify a weak out-of-plane ferroelectricity accompanying a stronger in-plane antiferro-orbital order, which in turn emerges under an even higher-energy antiferroelectric lattice structure.
Given the large energy scale of the antiferro-orbital order, the proposed mechanism naturally explains the puzzling high transition temperature of the FE phase in combination with a tiny electric polarization, while offering consistent explanations to many other known observations in this material.
More generally, this study reveals an unprecedented generic paradigm of electronic ferroelectricity that paves the way to new designs for ultrafast-switchable polarization ideal for next-generation non-volatile memory and other devices.

\textit{Effective $d^2$-Valence of W$^{4+}$ ions at Rydberg-Scale} -
We start by investigating the W valence using the density functional theory (DFT), known to give reliable charge distribution.
Given that the ionic charge is of Rydberg-scale, much stronger than that of the temperature scale ($<700$ K) of the observed lattice deformation~\cite{goldschmidt1926gesetze, sato2016extending}, the former must be insensitive to the latter.
Correspondingly, the lower-energy (lattice, orbital, and spin) physics~\cite{Dagotto2005}, such as the lattice transitions at 565 K~\cite{Tao_PhysRevB2020} and FE polarization at 350 K~\cite{Fei_nature2018}, must be under a fixed \textit{effective} ionic valence.
We thus purposely study the W ionic valence with a fictitious $1T$ structure of higher lattice symmetry~\ref{fig:d2_AFEO}(b), in which the W atoms are positioned at the center of the Te octahedra.

To decipher the emergent effective ionic valence and orbital structure, we applied the procedure of ``interaction annealing''~\cite{Ruoshi_annealArXiv_2024}.
At fictitiously large intra-atomic interaction $U>10$ eV, Fig.~\ref{fig:d2_AFEO}(a) shows that multiple ionic configurations are found stable, including $d^2$, $d^3$, and $d^4$.
Furthermore, upon slowly ``cooling'' the interaction $U$ to a realistic value (3 eV)~\cite{Kirchner2021, Linnartz2022}, the lowest-energy $d^2$ configuration is found to adiabatically flow to the ground state of the system.
In contrast, the $d^4$ and $d^3$ configurations become unstable below $U\sim 9.8$ eV and $U\sim 5.7$ eV, respectively.
Such an effective $d^2$ charge naturally explains the observed diamagnetic response~\cite{Yang2021}, when doubly occupying the orbital of the lowest energy.
That is, upon absorbing all quantum fluctuation, the dynamics of the emergent W ions are those of $d^2$-charged ions, corresponding to effective W$^{4+}$ ions.

\textit{Antiferroelectric Lattice Order at eV-Scale} -
Not surprisingly, the very small size of the effective $d^2$ W$^{4+}$ ions would promote further lattice deformation to the observed $T_d$ structure~\cite{goldschmidt1926gesetze, sato2016extending,shannon1976revised}.
Indeed, as shown in Fig.~\ref{fig:d2_AFEO}(b), similar to the typical FE BaTiO$_3$, a small ion in an octahedron would displace toward one of the triangular faces of the octahedron and in turn expand its area and the associated bond lengths.
For edge-sharing lattices as in WTe$_2$, such lattice displacements would have to be coordinated between neighboring octahedra [as indicated by the arrows in Fig.~\ref{fig:d2_AFEO}(b)] and thus result in the observed in-plane lattice structure.

\begin{table}[!t]
\caption{Estimated~\cite{supplementary} average deviations of W-Te bond length, $\Delta=\sum_{i=1}^6\left|d_i-d_{\text{mean}}\right|$, and the Te-W-Te intersection angle $\phi_i$, $\Sigma=\sum_{i=1}^{12}\left|90^\circ-\phi_i\right|$, of the local WTe$_6$ octahedron, using ionic radii of W$^{4+}$Te$^{2-}_2$ ($d^2$) and W$^{3+}$Te$^{1.5-}_2$ ($d^3$) configurations~\cite{shannon1976revised}.
Evidently, the $d^2$ configuration gives superior agreement with the experimental structure~\cite{Tao_PhysRevB2020}.}
\setlength{\tabcolsep}{8pt}
\vspace{5pt} 
\begin{tabular}{c|c|cc}
\toprule
\rule{0pt}{2.5ex} 
& \raisebox{0.2ex}{\begin{tabular}[c]{@{}c@{}}Exp.~\cite{Tao_PhysRevB2020}\end{tabular}} 
& \raisebox{0.2ex}{\begin{tabular}[c]{@{}c@{}}W$^{4+}$Te$^{2-}_2$ ($d^2$)\end{tabular}} 
& \raisebox{0.2ex}{\begin{tabular}[c]{@{}c@{}}W$^{3+}$Te$^{1.5-}_2$ ($d^3$)\end{tabular}} \\
\hline
\rule{0pt}{2.6ex}
$\Delta (\mathbf{\AA})$ & 0.306 & 0.319 & 0.073 \\
$\Sigma (^{\circ})$ & 145.200 & 162.963 & 204.611\\
\botrule
\end{tabular}
\vspace{-0.3cm}
\label{tab:ionic_radius}
\end{table}

Indeed, our DFT structural optimization with such effective $d^2$ W$^{4+}$ ions gives a distorted $T_d$ structure consistent with the experiment.
In fact, one can reach the same intuitive conclusion via simple ionic radius-based analysis of the local WTe$_6$ octahedron. 
Tab.~\ref{tab:ionic_radius} gives the calculated~\cite{supplementary} average deviations of both W-Te bond length, $\Delta=\sum_{i=1}^6\left|d_i-d_{\text{mean}}\right|$, and the Te-W-Te intersection angle $\phi_i$, $\Sigma=\sum_{i=1}^{12}\left|90^\circ-\phi_i\right|$, of the local WTe$_6$ octahedron.
Evidently, compared with W$^{3+}$Te$^{1.5-}_2$ ($d^3$), W$^{4+}$Te$^{2-}_2$ ($d^2$) gives a superior agreement with the experimental structure.
In fact, the negligible $\Delta$ for W$^{3+}$Te$^{1.5-}_2$ indicates that W$^{3+}$ ion is too large to be compatible with the $T_d$ structure. 

\begin{figure}
\centering
\includegraphics[width=\columnwidth]{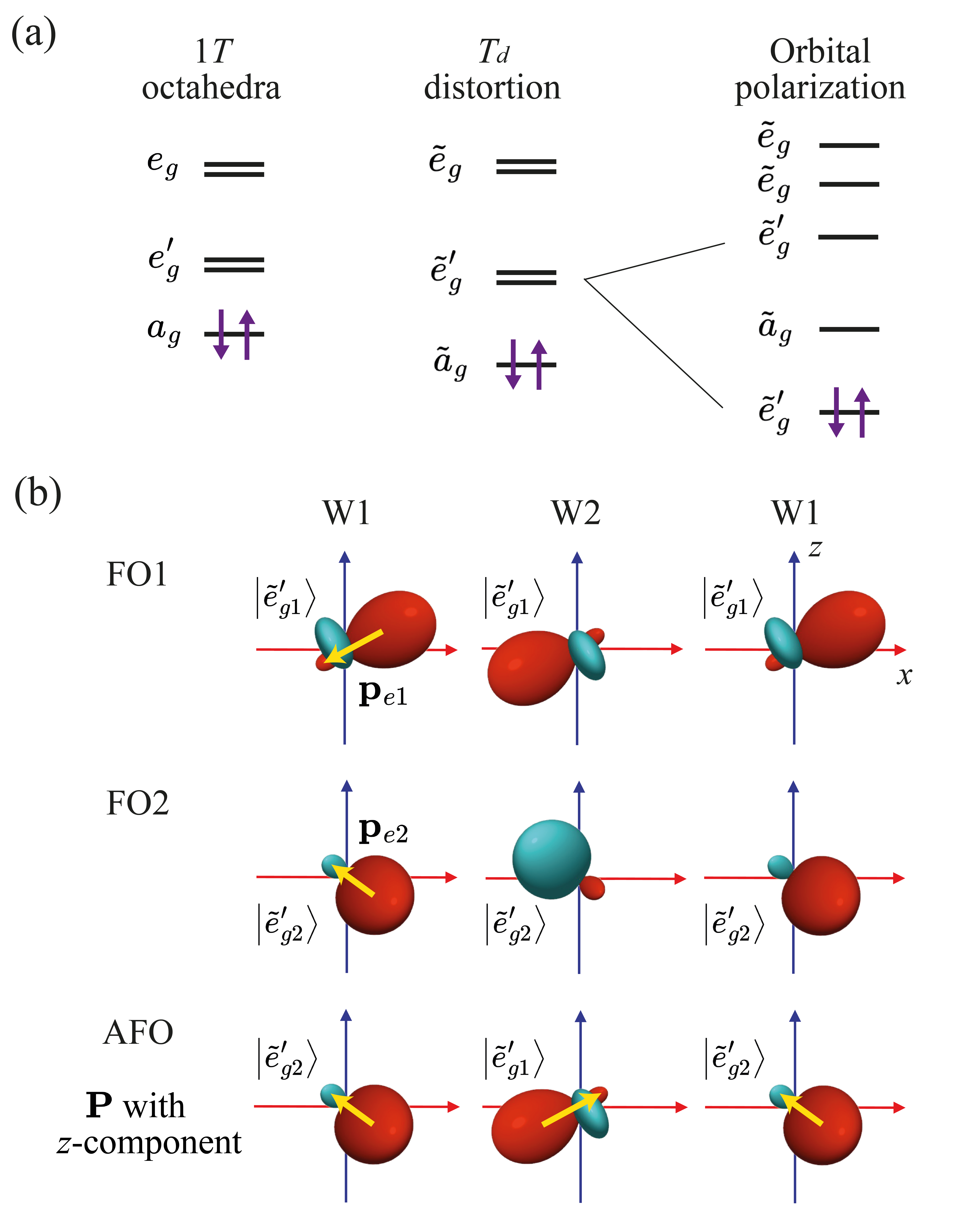}
\vspace{-0.7cm}
\caption{Illustration of the emergence of orbital polarization and the corresponding ferro-orbital orders.
(a) Local electronic structures of a W ion in the 1$T$ and the distorted $T_d$ structures, together with the orbital polarization that emerges in the latter.
(b) Exaggerated illustration of the polarized orbitals $|\tilde{e}_{g1}^\prime\rangle$ and $|\tilde{e}_{g2}^\prime\rangle$ in W ions, hosting dipolar moments $\mathbf{p}_{e1}$ and $\mathbf{p}_{e2}$ (yellow arrows), respectively, having opposite out-of-plane $z$-components.
In contrast to ferro-orbital orders (FO1 and FO2), the lower-energy antiferro-orbital order (AFO) produces a global polarization $\mathbf{P}$ along the $z$-direction.}
\label{fig:AFO}
\vspace{-0.5cm}
\end{figure}

Interestingly, in such $T_d$ lattices the coordinated lattice displacement corresponds to an antiferroelectric (AFE) order.
Indeed, Fig.~\ref{fig:d2_AFEO}(b) illustrates that the displacement of W ions breaks the local inversion symmetry, while the coordination between neighboring octahedra dictates anti-alignment of the local electric (red and blue) dipoles.
That is, despite the lack of FE component in this structure, large local electric dipole of density $\sim4.71~\mu\mathrm{C/cm^2}$ (one order bigger than the observed FE dipole $\sim0.27~\mu\mathrm{C/cm^2}$) are always present in all known experimental conditions~\cite{Lee_SciRep2015, Zhou_AIPadv2016, Lu_PhysRevB2016}, according to the eV-scaled energy difference in our lattice optimization.
Therefore, all physics below the eV-scale, such as the low-temperature FE phase below 350 K, must take place under the environment of this AFE order, as well as the effective $d^2$ valence of W.

\begin{figure}
\centering
\includegraphics[width=\columnwidth]{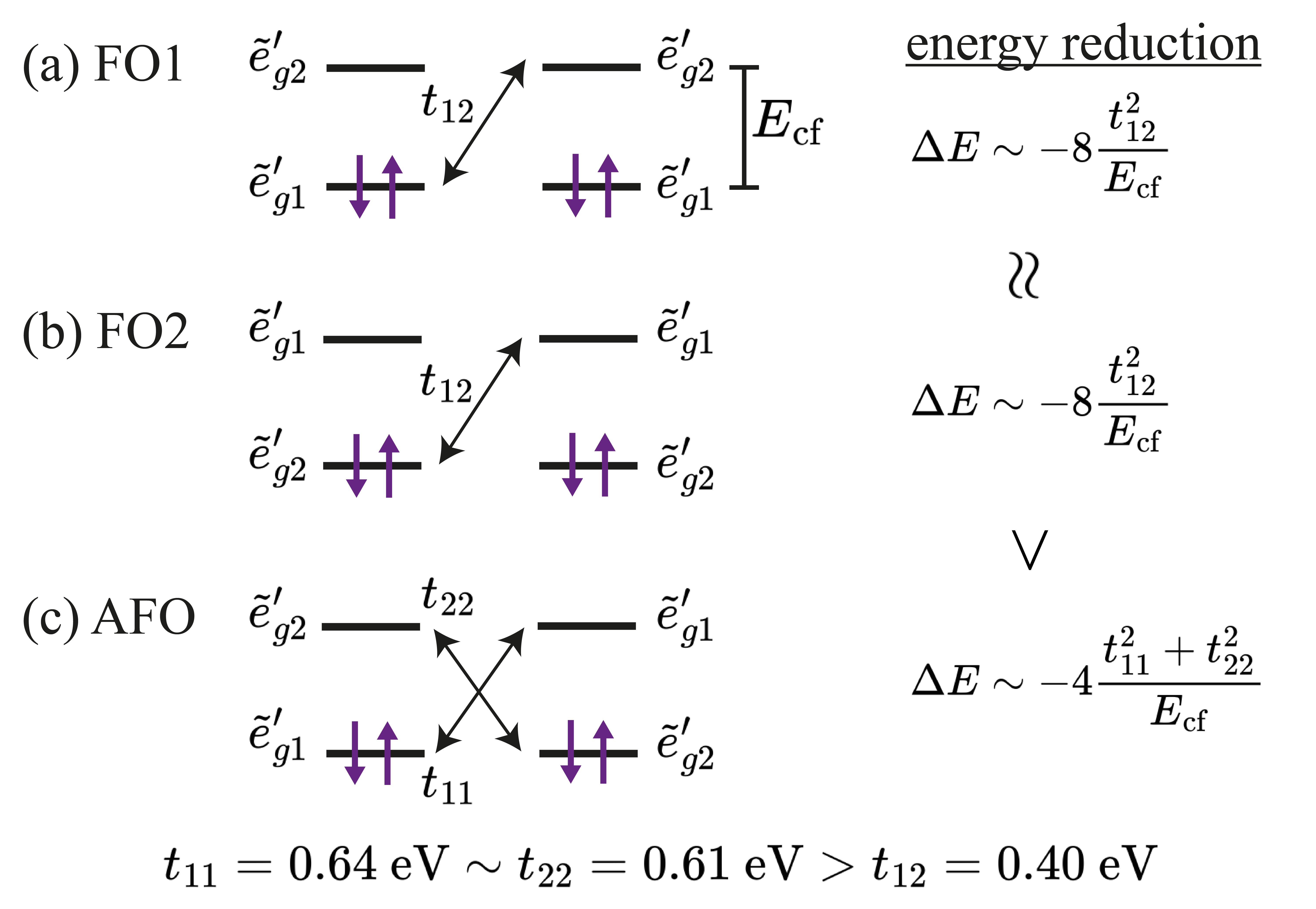}
\vspace{-0.5cm}
\caption{Illustration of the stronger energy reduction of the (c) AFO order through virtual kinetic processes against the charge fluctuation energy $E_\mathrm{cf}$, in comparison with the (a) FO1 and (b) FO2 orders, due to larger intra-orbital hopping strengths, $t_{11}$ and $t_{22}$, than the inter-orbital ones, $t_{12}$.
}
\label{fig:AFO_2}
\vspace{-0.5cm}
\end{figure}

\textit{Antiferro-orbital (AFO) order at sub-eV scale -}
Naturally, the lower lattice symmetry of the $T_d$ structure has a direct implication on the electronic orbitals relevant for lower-energy physics.
Specifically, the lack of parity around W ions implies hybridization between W-$d$ orbitals and W-$p$ orbitals, such that the low-energy effective $\tilde{d}$ orbitals contain finite dipole moments.
For an easier visualization, Fig.~\ref{fig:AFO}(b) exaggerates such hybridization in the effective $|\tilde{e}_{g1}^\prime\rangle\equiv\frac{1}{\sqrt{2}}(|e'_{g1}\rangle+\ket{p_x})$ and $|\tilde{e}^\prime_{g2}\rangle\equiv\frac{1}{\sqrt{2}}(|e'_{g2}\rangle-\ket{p_y})$ orbitals, both showing uneven spatial distribution reflecting their finite dipolar moments.
Notice that their corresponding dipolar moments are of different directions, with \textit{opposite} out-of-plane $z$-components.

Interestingly, the symmetry group within the subspace of the W-$d$ orbitals is not reduced by the lowered lattice symmetry.
This is evident in Fig.~\ref{fig:AFO}(a) from the preservation of degeneracy in the DFT energies of $e_g$ and $e_g^\prime$ Wannier orbitals~\cite{Ku2002}, when lowering the symmetry from 1$T$ to $T_d$ lattice.
Correspondingly, with two electrons fully occupying the lowest-energy $a_g$ orbital, there is \textit{a priori} no obvious tendency toward orbital polarization.

Nonetheless, at sub-eV scale our DFT calculation found lower total energy by more than 100 meV per formula unit, when the W$^{4+}$ ions develop orbital polarization by doubly occupying the $|\tilde{e}^\prime_{g1}\rangle$ or $|\tilde{e}^\prime_{g2}\rangle$ orbital instead of the $|\tilde{a}_g\rangle$ orbital.
Figure~\ref{fig:AFO}(b) gives three examples of various long-range orbital ordered states found in our calculation.
Essentially, the energy gain of the system by polarizing the $\tilde{e}_g^\prime$ orbitals exceeds the cost of crystal-field splitting between the $\tilde{e}_g^\prime$ and $\tilde{a}_g$ orbitals, as shown in Fig.~\ref{fig:AFO}(a).

Among the three states shown in Fig.~\ref{fig:AFO}(b), the antiferro-orbital (AFO) configuration turns out to be the most energetically favored.
This can be understood by the virtual kinetic energy gain $\sim 2\frac{t^2}{E_\mathrm{cf}}$ per W-W bond as shown in Fig.~\ref{fig:AFO_2}, where $t$ denotes the kinetic hopping strength and $E_\mathrm{cf}$ the charge fluctuation energy.
Since the intra-orbital hopping, $t_{22}=0.61 \ \mathrm{eV} \sim t_{11} = 0.64$ eV, is much larger than the inter-orbital one, $t_{12} =  t_{21} = 0.40$ eV, the virtual kinetic energy gain of the AFO state is therefore $(t_{11}^2+t_{22}^2)/(2t_{12}^2)\sim 3$ times stronger than that of the FO states.

\textit{Accompanying Ferroelectric Order} - Interestingly, this sub-eV scaled AFO state also hosts an accompanying out-of-plane ferroelectric order.
As illustrated in Fig.~\ref{fig:AFO}(b), in contrast to the fully compensated orbital dipole moments in the FO states, those in the AFO state host a global electric dipole $\mathbf{P}$ with an out-of-plane $z$-component.
Therefore, a long-range FE order would always accompany the AFO order.
In addition, the in-plane component of $\mathbf{P}$ cancels between the two layers in the $T_d$ structure, given their 180$^\circ$ rotated structure~\cite{Tao_PhysRevB2020, Zhou_AIPadv2016}.
Naturally, this AFO state and its symmetric counterpart, with interchanged $|\tilde{e}_{g}^\prime\rangle$ occupation and opposite $\mathbf{P}$, are energetically degenerate and switchable via external electric field.

\begin{figure}
\centering
\includegraphics[width=\columnwidth]{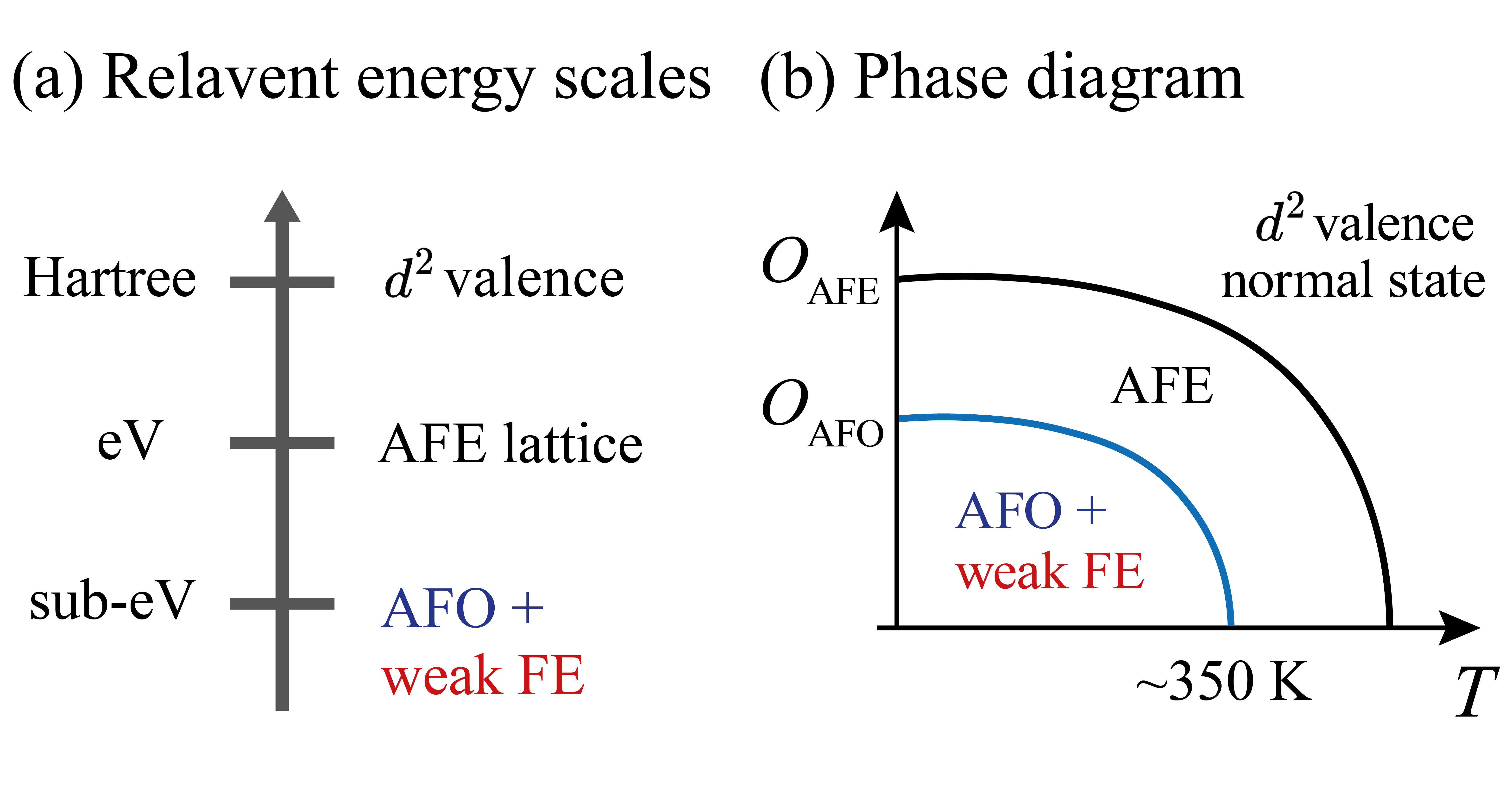}
\vspace{-0.8cm}
\caption{Relevant energy scales of dominant physics and the corresponding phase diagram.
(a) Relevant energy scales for 1) the effective $d^2$ valence at Hartree, 2) the antiferroelectric (AFE) lattice at eV, and 3) at sub-eV the antiferro-orbital (AFO) order, with which the weak ferroelectric (FE) order accompanies.
(b) Schematic phase diagram containing three phases: 1) the normal state with $d^2$ valence at the highest temperature, 2) the AFE phase at lower temperature, and 3) the AFO phase accompanied by a weak FE order.
For the latter two phases with spontaneously broken symmetries, the corresponding order parameters grow at lower temperature.}
\label{fig:phase_diagram}
\vspace{-0.6cm}
\end{figure}

This mechanism of AFO order accompanying FE order offers a perfect explanation for the puzzling high FE transition temperature with a tiny FE polarization.
On the one hand, the transition temperature is that of the sub-eV scaled AFO order and thus easily survives thermal fluctuation at 350 K.
On the other, the local \textit{electronic} dipoles result from the broken local parity-induced $d$-$p$ hybridization and can thus be very small.
Indeed, along the out-of-plane direction, our DFT obtained $\tilde{e}^\prime_{g1}$ and $\tilde{e}^\prime_{g2}$ orbitals give $\mathbf{p}_{e, z}$ components of approximately 0.188 $\mu$C/cm$^2$ and 0.085 $\mu$C/cm$^2$, respectively~\cite{supplementary}, which are more than \textit{two orders} of magnitude smaller than the typical local dipoles in FE perovskites.
This tiny polarization density is in reasonable agreement with the experimental observation~\cite{Fei_nature2018}, considering the unavoidable long-range quantum fluctuation present in real materials.

As a matter of fact, among all common electronic orders, orbital orders are the special ones more likely to produce the observed hysteresis in conductance measurements below 350 K ~\cite{Fei_nature2018, Sharma_sciadv2019, Xiao2020_natphy}.
Unlike most electronic orders that spontaneously break \textit{continuous} symmetries, orbital orders, like the FE order, typically break \textit{discrete} symmetries, due to orbitals' strong coupling to the atomic lattice with discrete point-group symmetries.
Similarly, the resulting \textit{anisotropy} of `pseudo-spins'~\cite{kanamori_crystal_1960} for the orbital polarization naturally explains why such order can sustain the long-range thermal fluctuation~\cite{PhysRev.65.117} that destroys strict ordering of all continuous symmetries in two dimension~\cite{Mermin-Wagner_1966}.
In fact, the expected reduction of thermal fluctuation in samples with increased layers is in perfect agreement with the enhancement of observed hysteresis~\cite{Fei_nature2018}.
(Depending on the sample homogeneity and the quantitative strength of long-range fluctuation, monolayer materials can either display a weak order~\cite{Yuan_NatComm2019} or completely lose the order~\cite{Fei_nature2018}.)

\textit{Additional consistency with experimental observations} - Contrary to the serious inconsistencies between the experimental observations and the popular proposal of interlayer sliding of atomic position, this electronic mechanism is naturally resolve these inconsistencies.
First, as illustrated in Fig.~\ref{fig:phase_diagram}, such \textit{electronic} mechanism does not require a higher \textit{structural} symmetry above the $\sim$350 K transition temperature, just as observed experimentally~\cite{Tao_PhysRevB2020, Zhou_AIPadv2016}.
Second, it generates an \textit{additional} electronic ordering on top of the already broken inversion symmetry, as evident from the observed strong asymmetry in the hysteresis curves~\cite{Fei_nature2018, Sharma_sciadv2019, Xiao2020_natphy} slightly below the $\sim$350 K transition temperature.
Third, the much smaller electronic mass implies a much weaker switching $\mathbf{E}$-field, in agreement with experimental observations~\cite{Fei_nature2018}.

\begin{figure}
\centering
\includegraphics[width=0.9\columnwidth]{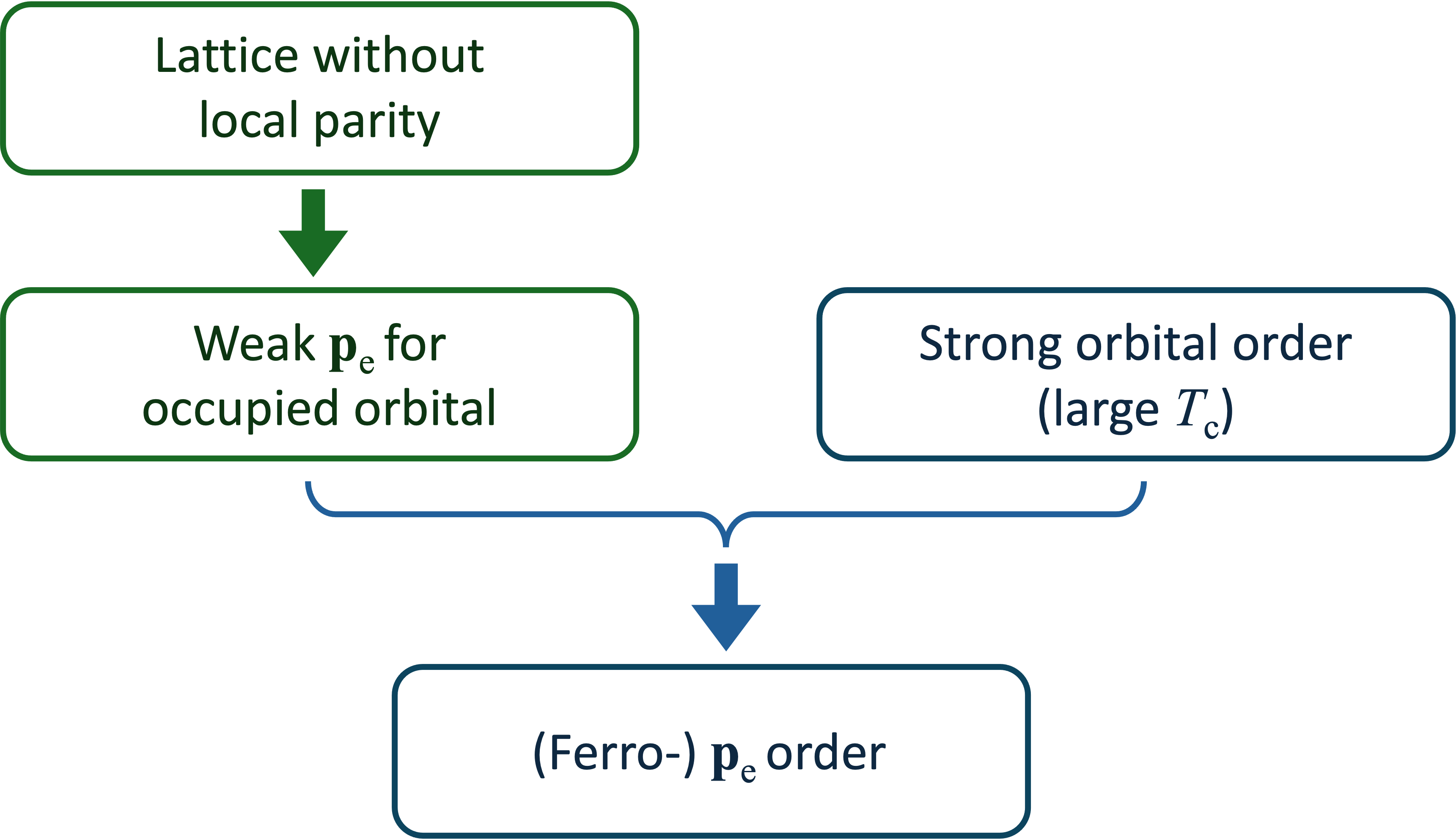}
\vspace{-0.2cm}
\caption{Schematic of a general \textit{electronic} mechanism for ferroelectricity with a small ordered moment but a high transition temperature.
In a lattice with local parity, local orbital polarization is accompanied by \textit{weak} electric dipole $\mathbf{p}_\mathrm{e}$ through orbital hybridization.
Upon a \textit{strong} long-range order of the orbitals below a high transition temperature $T_c$, ordering of the electric dipole follows.
}
\label{fig:overall_schematic}
\vspace{-0.5cm}
\end{figure}

\textit{General applicability - }
Our proposed electronic mechanism is quite general in its applicability.
In essence, as illustrated in Fig~\ref{fig:overall_schematic}, it only requires a lattice structure without local parity symmetry, to enable a weak electric dipolar moment in each electronic orbital through hybridization.
A strong long-range ordering of the orbitals would, therefore, be accompanied by the ordering of the weak electric dipole $\mathbf{p}_e$.
The simplicity of the above conditions should therefore be easily realized in many functional materials.

Typical characteristics of such ordering of \textit{electronic} dipoles include 1) lack of parity lowering in the atomic lattice 2) a rather high ordering temperature, 3) a comparably small dipolar order parameter, 4) hysteresis under switching processes, and 5) ultrafast switching with low energy cost.
Interesting, many of these characteristics have been observed in several novel functional materials.
To give a couple of examples, trilayer ReSe$_2$~\cite{xue_emergence_2025} and monolayer $d$1$T$-MoTe$_2$~\cite{Yuan_NatComm2019} both satisfy the conditions for our proposed mechanism, and furthermore display nearly all of these characteristics~\cite{Yuan_NatComm2019}.
(Obviously, the popular proposal of sliding layers is inapplicable to the monolayer $d$1$T$-MoTe$_2$.)
Therefore, it would be of great interest to further investigate these materials for direct experimental evidence of orbital ordering, for example, via electron microscopy~\cite{PhysRevR_2022_Peng} or electron energy loss spectroscopy~\cite{APL_iwashimizu_2021}.

In summary, in the representative 2D semi-metallic WTe$_2$, we identify a weak out-of-plane ferroelectricity accompanying a strong in-plane antiferro-orbital order, via a density-functional-based multi-energy-scale analysis of the system's broken symmetries.
This unusual low-energy correlation, which emerges from an antiferroelectric structure formed at much higher energy, naturally explains the puzzling observation of a high transition temperature despite the small magnitude of ordered polarization.
Our results reveal an unprecedented paradigm of \textit{electronic} ferroelectricity, broadly applicable to 2D polar metals, featuring ultrafast-switchable polarization ideal for next-generation non-volatile memory and other devices.


We acknowledge helpful discussions with Shuai Dong, Andrew M. Rappe, Jinning Hou, and Anthony Charles Hegg.
This work is supported by the National Natural Science Foundation of China (NSFC) \#12274287 and \#12042507, and the Innovation Program for Quantum Science and Technology (Project number: 2021ZD0301900).
F.G. also acknowledges the support from the International Postdoctoral Exchange Fellowship Program (YJ20210137) by the Office of China Postdoc Council (OCPC).
R.J. also acknowledges the support from UKRI Future Leaders Fellowship [MR/V023926/1]. 



\bibliographystyle{apsrev4-2}
\bibliography{2d_FE}

\end{document}